# On quantitative analysis of interband recombination dynamics: Theory and application to bulk ZnO

S. Lettieri,[1] V. Capello,[2] L. Santamaria,[2,3] and P. Maddalena[1,2]

[1]*Institute for Superconductors, Oxides and Innovative Materials, National Research Council (CNR-SPIN), U.O.S. Napoli, Via Cintia, I-80126 Napoli (Italy)*
[2]*Physics Department, University of Naples "Federico II", Via Cintia I-80126 Napoli (Italy)*
[3]*INO-CNR, Istituto Nazionale di Ottica, Sezione di Napoli, Via Campi Flegrei 34, I-80078 Pozzuoli (NA) Italy*

The issue of the quantitative analysis of time-resolved photoluminescence experiments is addressed by developing and describing two approaches for determination of unimolecular lifetime, bimolecular recombination coefficient, and equilibrium free-carrier concentration, based on a quite general second-order expression of the electron-hole recombination rate. Application to the case of band-edge emission of ZnO single crystals is reported, evidencing the signature of sub-nanosecond second-order recombination dynamics for optical transitions close to the interband excitation edge. The resulting findings are in good agreement with the model prediction and further confirm the presence, formerly evidenced in literature by non-optical methods, of near-surface conductive layers in ZnO crystals with sheet charge densities of about $3 \div 5 \ast 10^{13}$ cm$^{-2}$.

Photoluminescence (PL) spectroscopy is an optical characterization technique commonly used to probe bandgap energies and defect-related levels, identify impurities and provide information on the crystal quality and surfaces/interfaces properties.[1,2] In time-resolved variant (TRPL), the intensity of the PL emission induced by an instantaneous optical excitation source is detected as a function of time after excitation to probe the recombination dynamics of excited states. TRPL analysis provides qualitative and phenomenological information (e.g. exponential lifetimes) in a very direct manner. However, the extraction of *quantitative* information requires a proper modeling of the active processes and a congruent procedure for data analysis complementing the experimental results.

Going beyond frequently-used phenomenological fitting of the TRPL data and addressing the issue of quantitative determination of *actual* and non-phenomenological physical parameters characterizing the recombination dynamic represent the main motivations of the present study. We have dealt here with the quite general case of a "second-order" electron-hole recombination rate W, (i.e. a W depending on the excess charge density rho up to its quadratic term[3]) and, in particular, we propose and describe two data analysis approaches and experimentally test them in a specific case. For the latter, we focused our attention on zinc oxide (ZnO), in view of its importance as blue-UV light emitter[4-6] and optical gain medium for standard and/or random lasing.[7-10] Room-temperature PL investigations of bulk ZnO crystals evidences different kinetics for spectral regions positioned at photon energies below and above the PL peak: while the former is satisfactorily described by an exponential kinetics, a different kinetic regime is observed for optical transitions positioned above the excitonic peak.[11] The latter kinetics cannot be successfully interpreted in the framework of a multi-exponential decay dynamics, while we show here that it indeed exhibits the features characterizing a second-order recombination rate. The TRPL analysis approaches next described have thus been used, obtaining mutually consistent values for the recombination parameters and confirming the presence of ZnO conducting surface layers[12] with sheet carrier densities in accordance with literature findings.

We start our discussion by considering an n-type direct-gap semiconductor, indicating by $N$ and $\rho$ the equilibrium free carrier density in conduction band and the non-equilibrium (time dependent) density of excess carriers created by optical excitation. After pulsed laser excitation, the system recovers to equilibrium through recombination processes, whose time evolution will be dictated by the rate equation d$\rho$/dt = W = W$_R$+W$_{NR}$, where W, W$_R$ and W$_{NR}$ represent the total, the radiative and the non-radiative recombination rate respectively. As previously mentioned, we consider a W($\rho$) expansion up to the second order term, as W($\rho$) = $\rho/\tau$ + b$\rho^2$, where $\tau$ and b are the unimolecular recombination lifetime and the bimolecular coefficient, respectively.

By solving the rate equation d$\rho$/dt=W($\rho$(t)), the following decay kinetics is obtained for the excess carrier density:

$$\rho(t) = \frac{\rho_0 \exp(-t/\tau)}{1+\tau b \rho_0 \left[1-\exp(-t/\tau)\right]} \qquad (1)$$

where $\rho_0 = \alpha F/\hbar\omega = \gamma F$ is the initial (t = 0) density of photo-generated free carriers, $\alpha$ is the absorption coefficient at excitation photon energy $\hbar\omega$, F is the excitation fluence (i.e. the energy absorbed by the material per unit area)[13] and where the constant $\gamma = \alpha/\hbar\omega$ is defined for convenience.

The time-dependent PL yield $\phi$(t) is proportional to the radiative rate W$_R$ (emitted photons flux), whose expression for electron-hole pair recombination is W$_R$ = Bnp = BN$\rho$ + B$\rho^2$ (n=N+$\rho$, p=$\rho$) and where B is the radiative bimolecular coefficient.[14] Using Eq.(1.1) in $W_R(\rho(t))$, the following time evolution of TRPL signal is obtained (apart from a proportionality factor that is unessential for data analysis):[15]

$$\phi(t) \propto \frac{\rho_0 \exp(-t/\tau)}{1+b\tau\rho_0[1-\exp(-t/\tau)]} + \frac{B\rho_0^2 \exp(-2t/\tau)}{\{1+b\tau\rho_0[1-\exp(-t/\tau)]\}^2} \quad (2)$$

Inspecting the rate equation $d\rho/dt = \rho/\tau + b\rho^2$, it is immediately seen that the quadratic term $b\rho^2$ in the rate equation is dominant as long $\rho(t) > \tau b^{-1}$. As the maximum carrier density $\rho_0$ is achieved at the beginning of the decay process and as $\rho_0 \propto F$, the time behavior of the $\phi(t)$ function is characterized by a non-exponential initial slope, followed by a (fluence-independent) exponential slope ($\phi(t) \sim \exp(-t/\tau)$ for $t \gg \tau$).

Calculation of $\phi(t)$ curves are reported in Fig. 1, in order to evidence the evolution from a single-exponential behavior toward the "second-order" kinetics as the density of excited carriers is increased. As shown later, the same behavior is indeed exhibited by the investigated "high-energy branch" of ZnO PL.

For the present work, room temperature TRPL measurements were carried out on two (0001)-oriented single crystal ZnO substrates (CrysTec GmbH, hydrothermal growth, 500 µm thickness) coming from the same batch and both annealed for 30 min in nitrogen (T=900° C). Experiments were performed by means of a setup including a frequency-tripled Nd:YAG laser as excitation source (355 nm wavelength, 10 Hz repetition rate, 20 ps time duration of laser pulses) and a 250 mm focal length spectrometer coupled with a single-sweep streak camera for detection, allowing both time-resolved and wavelength-resolved PL measurement ("TRPL maps"), as represented in Fig. 2.

The TRPL maps were acquired for values of excitation fluence F ranging from 0 to about 200 µJcm$^{-2}$. Neither stimulated emission nor electron-hole plasma emission were observed in our measurements within the investigated fluence range.[16] In other words, no dynamic (time-dependent) renormalization of transition energies has been induced within the time range (700-800 ps typically) investigated in our measurement.

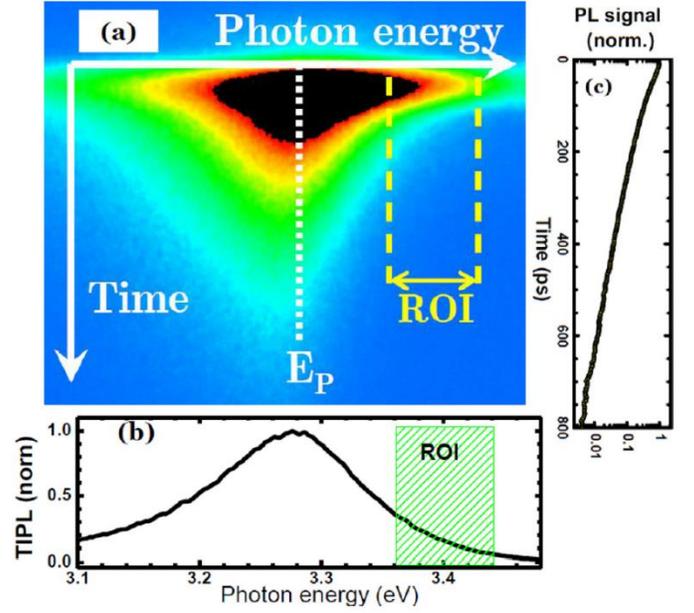

FIG. 2. (a): Representative PL intensity map of ZnO vs. photon energy (horizontal coordinate) and time (vertical coordinate). The total vertical extension corresponds to about 600 ps. (b): Time-integrated PL (TIPL) spectrum. (c): PL decay profile obtained by spectral integration over the ROI.

As already mentioned previously, we focused our study on a spectral "region of interest" (ROI, see Fig. 2) positioned at transition photon energies higher than the PL peak one. More precisely, we defined the ROI as an 80 meV wide energy interval whose center is shifted of about 120 meV respect to the PL peak

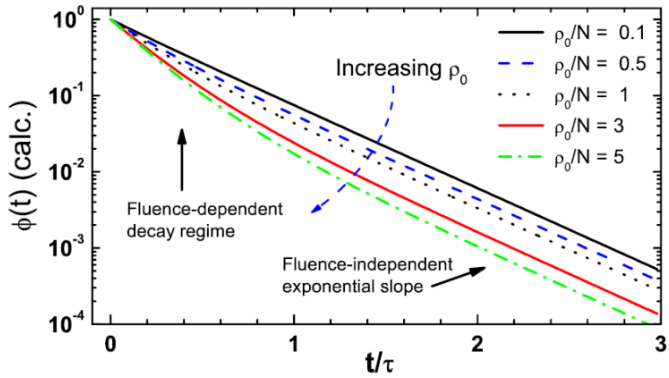

FIG. 1. Plots of calculated peak-normalized $\phi(t)$ function (Eq. (2)) for different values of $\rho_0/N$. The following values have been used for the calculations: $\tau = 0.5$ ns, $N = 10^{19}$ cm$^{-3}$, $b = 10^{-10}$ cm$^3$s$^{-1}$.

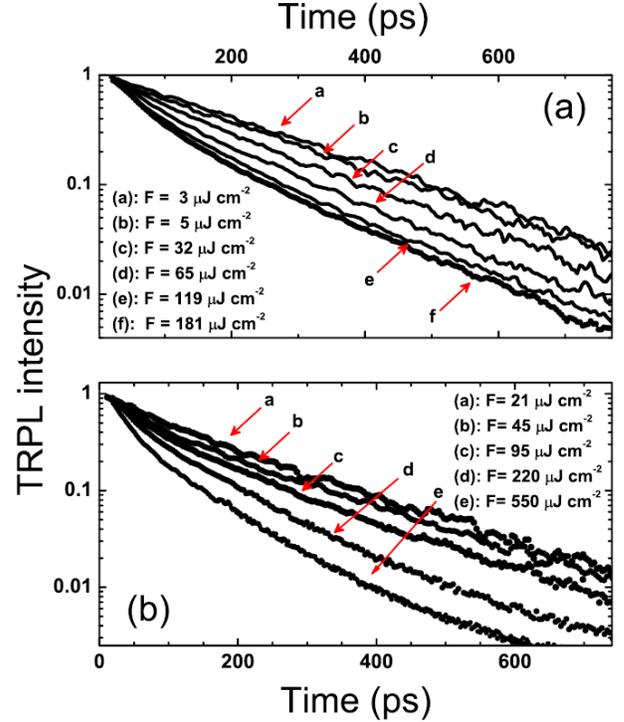

FIG. 3. Peak-normalized experimental TRPL curves for the two investigated ZnO samples A (top) and B (bottom) obtained at increasing excitation fluence F.

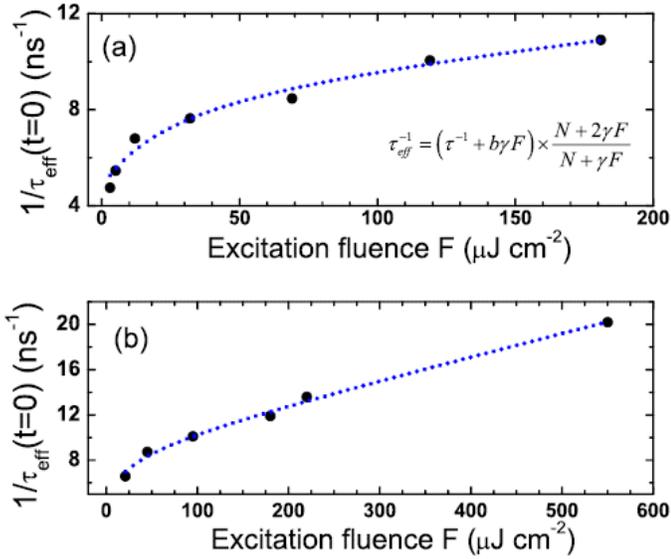

FIG. 4. Values for $1/\tau_{eff}(t=0)$ expressed in ns$^{-1}$ units vs. excitation fluence for the investigated samples A (top) and B (bottom). Dotted lines are the best-fit curves obtained using Eq. (3), which is also reported in inset of top panel for reader's convenience.

($E_P$) and obtained the TRPL curves by integration of the PL intensity over the ROI.[17]

Peak-normalized TRPL curves of the two investigated ZnO samples acquired at different values of F are reported in Fig. 3, remarkably evidencing the features that characterize the second-order recombination.

In particular, as the excitation fluence was increased the PL kinetics underwent a transition from a nearly single-exponential behavior (low fluences) to a different kinetics characterized by an initial non-exponential decay with fluence-dependent slope plus a following (fluence-independent) exponential decay. This encourages the idea to develop data analyses funded on the use of Eq. (2).

We first show a "fast" procedure for extracting the relevant physical parameters and its application to our measurements. The method relies on extracting the instantaneous "effective lifetime" $\tau_{eff}$, defined by the position $\tau_{eff}^{-1} = -(d\phi/dt)/\phi(t)$. Using the expression of radiative recombination rate and after a few simple algebra the following expression can be obtained for the initial (t=0) effective lifetime:

$$\tau_{eff}^{-1} = \left(\tau^{-1} + b\rho_0\right) \cdot \frac{N + 2\rho_0}{N + \rho_0} \quad (3)$$

The above relation expresses an initial logarithmic slope $1/\tau_{eff}(t=0)$ whose lower value is limited by the inverse of unimolecular lifetime in the low fluence limit (i.e. $\tau_{eff}(0) \cong \tau$ in the limit F → 0) and increasing as a function of $\rho_0$. Taking into account Eq. (3), the effective lifetime can be obtained from $\phi(t)$ by numerically calculating its time derivative $d\phi/dt$ and by extracting the value of $(d\phi/dt)/\phi(t)$ at t = 0. Next, values of b, $\tau$ and N can be obtained from data fitting using Eq. (3) in which the excitation fluence is introduced through $\rho_0 = \gamma F$.

Here, a reference value of $\alpha = 1.5*10^5$ cm$^{-1}$ for the ZnO linear absorption coefficient at 355 nm wavelength is used,[18] giving the $\tau_{eff}^{-1}$ values vs. F reported in Fig. (4).

The best-fit curves obtained by means of Eq. (3) are plotted as dashed lines. A satisfactorily consistence between the experimental behavior and theoretical predictions is evidenced, further supported by the fact that the limit value for $\tau_{eff}^{-1}(0)$ at nominal zero fluence tends to the fluence-independent logarithmic slope $\tau$ of TRPL curves.

Values for N, b and $\tau$ obtained through the best-fit of experimental data are reported in Table 1 and discussed later.

The above described $\tau_{eff}$ approach provides easy and fast estimation of the recombination parameters, while having obvious drawbacks (limited statistics, no use of the entire PL time evolution). Furthermore, it does not provide a strong test for the model we used to describe the recombination dynamics.

Therefore, we developed another approach, relying on the use of optical excitation fluence F as an additional degree of freedom for TRPL probing. The basic idea is to consider the $\phi(t)$ as a function of two variables, namely time (t) and excitation fluence F, the latter being proportional to the density of photo-generated carriers through $\rho_0 = \gamma F$. Thus, the whole set of TRPL curves obtained at different values of excitation fluence can be simultaneously fitted through Eq. (2), considering the bimolecular coefficient (b), the unimolecular lifetime ($\tau$) and the equilibrium charge carrier density (N) as best-fit parameters.

To this aim, we rearrange Eq. (2) obtaining the following function of two independent variables (x,y) and three parameters $\mu_i$ (i=1,2,3):

$$\Phi(x,y,\{\mu_i\}) = \frac{F(x,y,\{\mu_i\}) \cdot \left[1 + \mu_1 y F(x,y,\{\mu_i\})\right]}{1 + y\mu_1} \quad (4)$$

where the function F is defined as follows:

$$F(x,y,\{\mu_i\}) = \frac{\exp(-\mu_3 x)}{1 + y\mu_2\left[1 - \exp(-\mu_3 x)\right]}$$

and where a normalization factor respect to the x variable (i.e. $\Phi = 1$ for x = 0) has been introduced. The above function be used as global fitting function for the set of peak-normalized TRPL decays where (x,y) = (t,F), $\mu_1 = \gamma/N$, $\mu_2 = \gamma b\tau$ and $\mu_3 = 1/\tau$, allowing the determination of N, b and $\tau$.

The numerical procedure has been developed by writing a Fortran77 code based on MINUIT package. The results obtained by global fitting of the experimental TRPL curves are reported in Table 1, while the corresponding TRPL best-fit curves are reported in Fig.(5) for sample A.[19] As expectable due to the larger statistics, the global fitting parameters are affected by uncertainties that are significantly lesser than the ones obtained by effective lifetime procedure. It is however worth noting that all the values are consistent within their uncertainties.

It is interesting to discuss the N values reported in Table 1 by recalling that the optical excitation (and the subsequent PL emission) mainly involves a near-surface sheet, whose typical width is of the order of the optical penetration length $\Lambda = \alpha^{-1}$ (the value $\Lambda = 67$ nm is obtained from $\alpha = 1.5*10^5$ cm$^{-1}$).

Literature observation evidenced an almost ubiquitous presence of conductive surface layers in commercially available ZnO crystals, dominating the sample conductance at low temperatures and influencing the conductance even at room temperature.[20-22]

|   | Effective lifetime | Global fitting |
|---|---|---|
| A | $\tau = 210 \pm 3$ ps<br>$N = 5.1 \pm 1.3 * 10^{18}$ cm$^{-3}$<br>$b = 3.3 \pm 0.7 * 10^{-11}$ cm$^3$s$^{-1}$ | $\tau = 212 \pm 3$ ps<br>$N = 6.34 \pm 0.06 * 10^{18}$ cm$^{-3}$<br>$b = 3.19 \pm 0.06 * 10^{-11}$ cm$^3$s$^{-1}$ |
| B | $\tau = 203 \pm 3$ ps<br>$N = 5 \pm 2 * 10^{18}$ cm$^{-3}$<br>$b = 6.0 \pm 0.8 * 10^{-11}$ cm$^3$s$^{-1}$ | $\tau = 210 \pm 1$ ps<br>$N = 4.44 \pm 0.02 * 10^{18}$ cm$^{-3}$<br>$b = 5.36 \pm 0.07 * 10^{-11}$ cm$^3$s$^{-1}$ |

**TABLE 1:** Physical parameters of second-order recombination dynamics obtained by Effective Lifetime and Global Fitting approach for ZnO samples A and B.

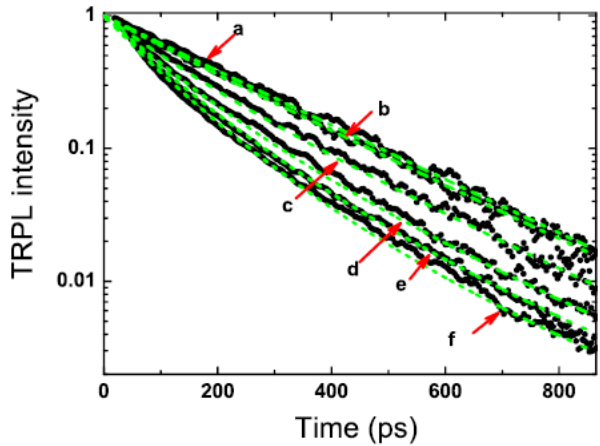

**FIG. 5**. Peak-normalized TRPL curves (sample A) and corresponding best-fit curves (green dashed lines) obtained by global-fitting procedure. See Fig. 3 caption for the correspondence between the curve letters and the excitation fluences).

Using temperature-dependent Hall effect and secondary-ion mass spectroscopy, Look and coworkers demonstrated that near-surface excess donor densities as high as $10^{19}$-$10^{20}$ cm$^{-3}$ can occur within the first 20 nm of ZnO sub-surface region, resulting from diffusion of group III atoms (Al, Ga, In) toward the surface of annealed samples.[22] A rough estimation of the sheet charge density s can be obtained in our case by $\sigma \cong N*\Lambda$: using the N values obtained by the global-fitting method, $\sigma \cong 4*10^{13}$ cm$^{-2}$ and $\sigma \cong 3*10^{13}$ cm$^{-2}$ for samples A and B (respectively) are obtained, in good accordance with $\sigma \cong 5.4*10^{13}$ cm$^{-2}$ reported by Look et al. for bulk ZnO single-crystals prepared in similar conditions.[22]

In summary, we addressed the issue of the quantitative determination of non-phenomenological parameters in TRPL experiments by working out two approaches allowing to obtain the basic quantities (namely unimolecular lifetime, bimolecular coefficient, equilibrium free-carrier density) characterizing the PL dynamics ruled by a second-order electron-hole recombination rate. As application, we considered the specific case of ZnO PL kinetics in the spectral region blue-shifted with respect to the dominant excitonic peak, where we evidenced a transition from single-exponential to second-order recombination kinetics, in agreement with the prediction for an interband charge carrier recombination. Application of the two TRPL analysis approaches here described gave mutually consistent results for the parameters and supports the presence of conducting surface layers, (previously evidenced in ZnO by means of other techniques) with sheet carrier density values in accordance with literature findings.